\documentclass[journal=undefined,manuscript=letter]{achemso}

\usepackage[T1]{fontenc} 
\usepackage{graphicx}
\usepackage{dcolumn}
\usepackage{bm}
\usepackage{mathrsfs}
\usepackage{amsmath}
\usepackage[linesnumbered,ruled,vlined]{algorithm2e}
\usepackage{float}
\usepackage{enumitem}
\usepackage{subfigure}
\usepackage{tabulary}
\usepackage{soul}
\usepackage{color}
\soulregister\cite7 
\soulregister\citep7 
\soulregister\citet7 
\soulregister\ref7 
\soulregister\pageref7 
\sethlcolor{white}

\newcommand{\norm}[1]{\left\lVert#1\right\rVert}
\newcommand{\abs}[1]{\left.|#1|\right.}

\newcommand{\bt}[1]{\bm{#1}}

\author{Yi Fan}
\affiliation{Hefei National Laboratory for Physical Sciences at the Microscale, University of Science and Technology of China, Hefei, Anhui 230026, China}
\author{Jie Liu}
\affiliation{Hefei National Laboratory for Physical Sciences at the Microscale, University of Science and Technology of China, Hefei, Anhui 230026, China}
\author{Zhenyu Li}
\email{zyli@ustc.edu.cn}
\affiliation{Hefei National Laboratory for Physical Sciences at the Microscale, University of Science and Technology of China, Hefei, Anhui 230026, China}
\author{Jinlong Yang}
\affiliation{Hefei National Laboratory for Physical Sciences at the Microscale, University of Science and Technology of China, Hefei, Anhui 230026, China}

\title{A Quantum Algorithm to Calculate Band Structure at the EOM Level of Theory}


\begin{document}
	\maketitle

	\begin{abstract}
		Band structure is a cornerstone to understand electronic properties of materials. Accurate band structure calculations using a high-level quantum-chemistry theory can be computationally very expensive. It is promising to speed up such calculations with a quantum computer. In this study, we present a quantum algorithm for band structure calculation based on the equation-of-motion (EOM) theory. First, we introduce a new variational quantum eigensolver algorithm named ADAPT-C for ground-state quantum simulation of solids, where the wave function is built adaptively from a complete set of anti-Hermitian operators. Then, on top of the ADAPT-C ground state, quasiparticle energies and the band structure can be calculated using the EOM theory in a quantum-subspace-expansion (QSE) style, where the projected excitation operators guarantee that the killer condition is satisfied. As a proof of principle, such an EOM-ADAPT-C protocol is used to calculate the band structures of silicon and diamond using a quantum computer simulator.
	\end{abstract}

	Many properties of a chemical system can be predicted by the electronic structure theory. Over the past decades, density functional theory (DFT)\cite{Hon64, Kohn65, Kohn96, CohMorYan12} has become one of the most widely used electronic structure method. However, its accuracy is difficult to be systematically improved since the  exchange-correlation part of the exact energy functional in DFT is unknown. For band structure calculation, GW approximation\cite{KliJirKal14} can be used to improve the accuracy of DFT results with local functionals. However, GW calculation usually has a high computational cost and its error is also hard to assess due to different self-consistency protocols adopted in practical GW calculations\cite{stan09}.
	
	Wave function based methods, such as configuration interaction (CI), M\o{}ller-Plesset perturbation theory, and coupled cluster theory (CC), offer a systematic way to approach the exact solution of the Schr\"odinger equation. For example, although typical CC calculations are truncated at single and double excitations (CCSD), we can in principle obtain accurate wave function by including all possible excitations. Recently, equation-of-motion coupled-cluster (EOM-CC) theory has been implemented in studying excitations and quasiparticle band structures in solid systems\cite{McCSunCha17, DittIzs19, WangBer20, GalloHum21, HamesJoh16}. Numerical calculations indicate that EOM-CCSD produces quasiparticle energies which are usually more accurate than those from the GW approximation\cite{LanBer18}. It can be proved that the EOM-CCSDT Green's function includes all diagrams contained in the GW approximation, along with many high-order vertex corrections. Although EOM-CC can reach a very high accuracy by incorporating high-order terms, \hl{it must be addressed that the computational cost also grows fast and quickly goes beyond the capability of current supercomputers. }
	
	\hl{Recent advances in quantum computing provides possibilities to obtain the exact solution to the many-electron Schr\"odinger equation within a polynomial time complexity.} Quantum algorithms for electronic structure calculations include quantum phase estimation (QPE) and variational quantum eigensolver (VQE)\cite{BraKit02, McAEndAsp20, CaoRomOls19, Pre18, GeoAshNor14, AspDutLov05, Wang08, PerMcCSha14, HemMaiRom18, NamChen20, SheZhaZha17, MalBabKiv16, KanMezTem17, ColRamDah18, McCRomBab16, LanWhi10, RomBabMcC18, vqe-excited-vqd, Mcclean17qse, YungCas2014}. The former can efficiently calculate the energy spectrum of a specific Hamiltonian. However, the high gate fidelity and error-correction devices required by QPE is typically far beyond the capability of the near-term noisy intermediate-scale quantum (NISQ) computers. By embedding efficient quantum state preparation and measurement in a classical optimization process, VQE yields a much shallower quantum circuit compared to QPE and it is thus more suitable for NISQ devices. The first experiment of VQE to study chemical systems was carried out on a photonic quantum processor to study the ground state of HeH\textsuperscript{+} \cite{PerMcCSha14}. In recent years, VQE has been demonstrated on most of the leading quantum computer architectures such as trapped-ion system\cite{HemMaiRom18, NamChen20, SheZhaZha17} and superconducting-qubit-based platform\cite{MalBabKiv16, KanMezTem17, ColRamDah18}. 
	
	The key ingredient of VQE is a parametric wave function ans\"atze. For example, one possibility is assuming that  the wave function can be written in a unitary coupled-cluster (UCC) form\cite{Kut82, BarKucNog89, TauBar06}. Recently, a more efficient and accurate wave function ans\"atze, adaptive derivative-assembled pseudo-trotter (ADAPT)\cite{GriEcoBar19}, is proposed, which relies on an iterative construction of the wave function. ADAPT algorithm provides a very accurate solution to the ground-state wave function, making it a favorable reference state for studying excitations under the EOM framework. Although VQE has been used to extract the bands of silicon from a tight-binding Hamiltonian\cite{CerShe20}, quantum algorithms for band structure calculations at a high ab-initio level is not yet available. It is thus desirable to extend the EOM theory to the ADAPT wave function and predict band structures in an EOM-ADAPT way on a quantum computer.
	
	Notice that the development of an EOM-ADAPT quantum algorithm for band structure calculation is not straightforward. First, like most quantum algorithms used in chemical simulations, the ADAPT algorithm is designed for molecular systems. In a previous study\cite{LiuWan20}, we have pointed out that a direct application of the ADAPT algorithm to periodic solid systems suffers from a residual error introduced by complex wave functions. A simple way to solve this problem is transforming complex Hartree-Fock orbitals at sampling $k$-points in a unit cell to real orbitals at the $\Gamma$ point in a supercell, which is termed as the K2G transformation\cite{LiuWan20}. However, this leads to a significant growth in the number of operator terms since the restriction of crystal momentum conservation is removed. At the same time, it is not convenient for band structure calculations especially when a non-orthogonal $k$-grid is used. Another difficulty in EOM-ADAPT band structure calculation is that the Hamiltonian similarity transformation technique which is widely used in EOM-CCSD \cite{szekeres2001on, Lev2004} to satisfy the killer condition is not applicable as detailed below. Therefore, extra efforts are required to avoid deviations between different EOM functionals, which can otherwise lead to additional errors in the predicted band gap.
	
	In this study, we present an EOM algorithm for quasiparticle band structure calculations on quantum computers. First, we introduce a modified ADAPT-VQE ans\"atze termed ADAPT-C for preparing the accurate ground state. In contrast to the K2G transformation scheme, the Hartree-Fock orbitals at sampling $k$-points are explicitly treated and the residual errors are minimized by introducing a complementary operator pool. Then, the quasiparticle states are obtained through a quantum subspace expansion (QSE)\cite{Mcclean17qse} style technique derived from the projected-operator EOM-IP and EOM-EA formalism\cite{szekeres2001on}. The converged ground-state wave function obtained with ADAPT-C is served as the reference state, and the excitation operators are transformed using the ground-state projector to satisfy the killer condition. Finally, the quasiparticle bands are extracted by identifying correct quasiparticle excitations from the solved eigenstates. We name this method as EOM-ADAPT-C, which, as a demonstration, is successfully applied to simulate band structures of silicon and diamond.
	
	\textbf{ADAPT-C algorithm.} Given the Hartree-Fock orbitals as linear combinations of Bloch atomic orbitals at sampling ${k}$ points, the Hamiltonian can be written in the second quantized form as
	\begin{equation}
		\begin{aligned}
			\label{eq-ham-pbc}
			\hat{H}&=\sum_{\bm{k}_p,\bm{k}_q}^{\prime}\sum_{p,q}h_{q\bm{k}_q}^{p\bt{k}_p}\hat{T}_{q\bm{k}_q}^{p\bt{k}_p} \\
			&+\frac{1}{2}\sum_{\substack{\bt{k}_{p},\bt{k}_{q}\\\bm{k}_{r},\bm{k}_{s}}}^{\prime}{\sum_{\substack{p,q\\r,s}}} g_{r\bm{k}_r s\bm{k}_s}^{p\bt{k}_p q\bt{k}_q} \hat{T}_{r\bm{k}_r s\bm{k}_s}^{p\bt{k}_p q\bt{k}_q}
		\end{aligned}
	\end{equation}
	with
	\begin{equation}
		\begin{aligned}
			\hat{T}_{q\bm{k}_q}^{p\bt{k}_p} &= \hat{a}^\dag_{p\bt{k}_p} \hat{a}_{q\bm{k}_q} \\
			\hat{T}_{r\bm{k}_r s\bm{k}_s}^{p\bt{k}_p q\bt{k}_q} &= \hat{a}_{p\bt{k}_p}^{\dagger} \hat{a}_{q\bt{k}_q}^{\dagger} \hat{a}_{r\bm{k}_r} \hat{a}_{s\bm{k}_s}
		\end{aligned}
	\end{equation}
	$h_{q\bm{k}_q}^{p\bt{k}_p}$ and $g_{r\bm{k}_r s\bm{k}_s}^{p\bt{k}_p q\bt{k}_q}$ are one- and two-electron integrals, respectively, which can be calculated with classical routines. Primed summations enforce crystal momentum conservation, i.e. the difference between summations of crystal momenta of creation and annihilation operators should be a reciprocal lattice vector. The creation and annihilation operators in Hamiltonian can be mapped to weighted Pauli strings and thus part of the quantum circuit using the Jordan-Wigner or Bravyi-Kitaev transformations\cite{JordanWigner28, BraKit02, SeeRich12, Tran18}. 
	
	The ground-state wave function can be obtained by solving the Schr\"odinger equation
	\begin{equation}
		\hat{H}|\Psi_0\rangle=E_{0}|\Psi_0\rangle
	\end{equation}
	via a variational algorithm, where the energy expectation value is minimized with a specific wave function ans\"atze. The UCC ans\"atze is a widely used physically motivated ans\"atze for quantum simulations, where the wave function takes the form
	\begin{equation}
		|\Psi\rangle=e^{\hat{T}-\hat{T}^{\dagger}}|\Phi_{0}\rangle
	\end{equation}
	$|\Phi_{0}\rangle$ is a reference state, typically the Hartree-Fock ground state. If truncated to the second order (UCCSD), the cluster operators are
	\begin{equation}
		\begin{aligned}
			\hat{T}&=\sum_{\bm{k}_i,\bm{k}_a}^{\prime}\sum_{a,i}\theta_{i\bm{k}_i}^{a\bt{k}_a}\hat{T}_{i\bm{k}_i}^{a\bt{k}_a} \\
			&+\frac{1}{4}\sum_{\substack{\bt{k}_{a},\bt{k}_{b}\\\bm{k}_{i},\bm{k}_{j}}}^{\prime}{\sum_{\substack{a,b\\i,j}}} \theta_{i\bm{k}_i j\bm{k}_j}^{a\bt{k}_a b\bt{k}_b} \hat{T}_{i\bm{k}_i j\bm{k}_j}^{a\bt{k}_a b\bt{k}_b}
		\end{aligned}
	\end{equation}
	It has been suggested that the performance of UCCSD can be improved by replacing occupied and virtual orbitals by general orbitals (UCCGSD)\cite{Maz04, MukKut04, Ron03, LeeHugHea19}. 
	\begin{equation}
		\begin{aligned}
			\hat{T}&=\frac{1}{2}\sum_{\bm{k}_p,\bm{k}_q}^{\prime}\sum_{p,q}\theta_{q\bm{k}_q}^{p\bt{k}_p}\hat{T}_{q\bm{k}}^{p\bt{k}} \\
			& +\frac{1}{4}\sum_{\substack{\bt{k}_{p},\bt{k}_{q}\\\bm{k}_{r},\bm{k}_{s}}}^{\prime}{\sum_{\substack{p,q\\r,s}}} \theta_{r\bm{k}_r s\bm{k}_s}^{p\bt{k}_p q\bt{k}_q} \hat{T}_{r\bm{k}_r s\bm{k}_s}^{p\bt{k}_p q\bt{k}_q}
		\end{aligned}
	\end{equation}
	Here, we use $\{i, j, k, \dots \}$, $\{a, b, c, \dots\}$, and $\{p, q, r, \dots\}$ to indicate occupied, virtual, and general orbitals, respectively.
	
	The exponential unitary operator in the UCC ans\"atze cannot be converted to a quantum circuit directly. In practice, it is decomposed into one- and two-qubit quantum gates using an approximated scheme named Trotter-Suzuki decomposition\cite{GriClaEco20, BabMcCWec15}:
	\begin{equation}
		e^{\hat{A}+\hat{B}}\approx(e^{\hat{A}/k}e^{\hat{B}/k})^{k}
	\end{equation}
	The corresponding UCC wave function has an expression of
	\begin{equation}
		|\Psi \rangle=\lim_{N\rightarrow\infty}{\prod_{k=1}^{N}{\prod_{i}^{N_{i}}{e^{\frac{\theta_{i}}{N}\hat{\tau}_{i}}|\Phi_{0} \rangle}}}
	\end{equation}
	with $\{\hat{\tau}_{i}\}$ being anti-Hermitian operators:
	\begin{equation}
		\hat{\tau}_{q\bm{k}_q}^{p\bm{k}_p}=\hat{T}_{q\bm{k}_q}^{p\bm{k}_p} - \hat{T}^{q\bm{k}_q}_{p\bm{k}_p}
	\end{equation}
	\begin{equation}
		\hat{\tau}_{r\bm{k}_r s\bm{k}_s}^{p\bm{k}_p q\bm{k}_q}=\hat{T}_{r\bm{k}_r  s\bm{k}_s}^{p\bm{k}_p q\bm{k}_q} - \hat{T}^{s\bm{k}_s r\bm{k}_r}_{q\bm{k}_q p\bm{k}_p}
	\end{equation}
	Here, general excitation operators $\{\hat{\tau}_{q\bm{k}_q}^{p\bm{k}_p},\hat{\tau}_{r\bm{k}_r s\bm{k}_s}^{p\bm{k}_p q\bm{k}_q}\}$ should also satisfy the crystal momentum conservation rule. The Trotterization procedure is to be truncated at finite order and the ordering of the Trotterized cluster operators is not unique, which may introduce a considerable Trotterization error\cite{GriClaEco20,BabMcCWec15}. 
	
	The recently proposed ADAPT ans\"atze\cite{GriEcoBar19} is capable of bypassing the Trotterization error by generating a compact sequence of unitary operators. Instead of optimizing a large number of UCC parameters simultaneously, a pseudo-Trotter wave function with finite variational parameters is constructed
	\begin{equation}
		|\Psi^{(k+1)} \rangle =e^{\theta^{(k+1)}\hat{\tau}^{(k+1)}} |\Psi^{(k)} \rangle
	\end{equation}
	where the exponential terms are taken from the operator pool $\mathcal{O}$ defined by anti-Hermitian operators $\hat{\tau_{i}} \in \mathcal{O} \equiv \{\hat{\tau}_{q\bm{k}_q}^{p\bm{k}_p},\hat{\tau}_{r\bm{k}_r s\bm{k}_s}^{p\bm{k}_p q\bm{k}_q}\}$.
	At the $k$th iteration, the energy is minimized through the variational procedure:
	\begin{equation}
		E^{(k)}=\min_{\{\vec{\theta}^{(l)}\}_{l=1}^{k}}{\{\langle \Psi^{(k)}|\hat{H}|\Psi^{(k)} \rangle \}}
	\end{equation}
	At the $(k+1)$th iteration, among all operators in $\mathcal{O}$, the one with the largest residual gradient is selected as $\hat{\tau}^{(k+1)}$. The residual gradient of operator $\hat{\tau}_{i}$ is defined as
	\begin{equation}
		\begin{aligned}
			R_{i}^{(k)}&=\left.\frac{\partial{E^{(k+1)}}}{\partial{\theta^{(k+1)}}}\right|_{\theta^{(k+1)}=0,\hat{\tau}^{(k+1)}=\hat{\tau}_{i}} \\
			&=\langle \Psi^{(k)} |[\hat{H},\hat{\tau}_{i}]| \Psi^{(k)}  \rangle
		\end{aligned}
	\end{equation}
	The $L2$-norm of residual gradients $\vec{R}^{(k)}=(R_{1}^{(k)},R_{2}^{(k)},\dots)$ is used as the convergence criterion
	\begin{equation}
		\label{eq-adapt-convg}
		\norm{\vec{R}^{(k)}}_{2}=\sqrt{\sum_{i}{\abs{R_{i}^{(k)}}^{2}}}<\varepsilon
	\end{equation}
	
	Such a convergence criterion guarantees that the real part of the anti-Hermitian contracted Schr\"odinger equation (ACSE)\cite{Maz07} is satisfied. The imaginary part of ACSE is automatically satisfied only for real-valued wave functions. Therefore, there is a residual error originated from the imaginary part of ACSE when the ADAPT algorithm is straightforwardly applied to solids with complex wave function\cite{LiuWan20}. In the UCC case, it is suggested to use complex cluster amplitudes to avoid this problem.\cite{Manrique21}
	
	In ADAPT, it is more natural to introduce a complementary operator pool $\mathcal{O}_{c}$
	\begin{equation}
		\mathcal{O}_{c}=\{\hat{\tilde{\tau}}_{q\bm{k}_q}^{p\bm{k}_p},\hat{\tilde{\tau}}_{r\bm{k}_r s\bm{k}_s}^{p\bm{k}_p q\bm{k}_q}\}
	\end{equation}
	where $\hat{\tilde{\tau}}_{q\bm{k}_q}^{p\bm{k}_p}$ and $\hat{\tilde{\tau}}_{r\bm{k}_r s\bm{k}_s}^{p\bm{k}_p q\bm{k}_q}$ are anti-Hermitian operators defined as
	\begin{equation}
		\hat{\tilde{\tau}}_{q\bm{k}_q}^{p\bm{k}_p}=i(\hat{T}_{q\bm{k}_q}^{p\bm{k}_p} + \hat{T}^{q\bm{k}_q}_{p\bm{k}_p})
	\end{equation}
	\begin{equation}
		\hat{\tilde{\tau}}_{r\bm{k}_r s\bm{k}_s}^{p\bm{k}_p q\bm{k}_q}=i(\hat{T}_{r\bm{k}_r s\bm{k}_s}^{p\bm{k}_p q\bm{k}_q} + \hat{T}^{s\bm{k}_s r\bm{k}_r}_{q\bm{k}_q p\bm{k}_p}).
	\end{equation}
	The wave function is then updated using the extended operator pool
	\begin{equation}
		\tilde{\mathcal{O}}= \mathcal{O} \cup \mathcal{O}_{c} =\{\hat{\tau}_{q\bm{k}_q}^{p\bm{k}_p},\hat{\tau}_{r\bm{k}_r s\bm{k}_s}^{p\bm{k}_p q\bm{k}_q},\hat{\tilde{\tau}}_{q\bm{k}_q}^{p\bm{k}_p},\hat{\tilde{\tau}}_{r\bm{k}_r s\bm{k}_s}^{p\bm{k}_p q\bm{k}_q}\}
	\end{equation}
	The size of the operator pool is doubled regardless of the Brillouin zone sampling scheme. At the same time, the crystal momentum conservation law is still preserved. We name such a modified ADAPT algorithm with a complementary operator pool as ADAPT-C. In ADAPT-C, variational parameters for operators in both $\mathcal{O}$ and $\mathcal{O}_c$ are kept to be real numbers.
	
	\textbf{EOM theory.}
	The EOM formalism based on the coupled-cluster theory\cite{eom-1, eom-2, eom-3, EOM-book} is a well-established approach to calculate excitation energies in quantum chemistry. In EOM theory, we focus on the excitation operator $\hat{\mathcal{R}}$ connecting the ground state $| \Psi_0 \rangle$ and a target excited state $|\Psi_{x}\rangle$. From the definition $\hat{\mathcal{R}} = | \Psi_{x} \rangle \langle  \Psi_0 |$, we can obtain an EOM-like equation for $\hat{\mathcal{R}}$
	\begin{equation}
		\label{eq-eom2}
		[\hat{H}, \hat{\mathcal{R}}]   = \Delta E_{x} \hat{\mathcal{R}} 
	\end{equation} 
	where the excitation energy is defined as $\Delta E_{x} = E_{x} - E_{0}$. Such an equation provides us a way to access excitation energies without the knowledge about the ground and excited states explicitly.
	
	Different types of target states can be studied in the EOM theory by using different excitation operators. Ionization potential (IP) and electron affinity (EA) can be obtained if the operator $\hat{\mathcal{R}}$ includes a different number of creation and annihilation operators and thus not electron conserving. The fundamental band gap of a $N$-electron neutral system is characterized as the energy difference between the IP and EA. Therefore, we can use the EOM-IP energy $\Delta E^{IP}_{x}$ and the EOM-EA energy $\Delta E^{EA}_{x}$ at sampling $k$-points to construct the valence and conduction bands.\cite{McCSunCha17} 
	
	Here, we truncate the EOM-IP and EOM-EA excitation operator up to the second order, which means the ionization energies are calculated by diagonalizing the Hamiltonian in the space of 1-hole (1$h$) and 2-hole, 1-particle (2$h$1$p$) states, while electron affinities are obtained in the space of 1-particle (1$p$) and 2-particle, 1-hole (2$p$1$h$) states. The corresponding excitation operators $\hat{\mathcal{R}}_{IP}$ and $\hat{\mathcal{R}}_{EA}$ are expressed as a linear combination of a set of Fermion excitation operators as
	\begin{equation}
		\begin{aligned}
			\hat{\mathcal{R}}_{IP}(\bm{k})&=\sum_{p}{r_{p\bm{k}}\hat{a}_{p\bm{k}}} 
			&+ \sum_{\substack{\bt{k}_p\\ \bm{k}_q, \bm{k}_s}}^{\prime}{\sum_{p, q, s}{r_{q\bm{k}_q s\bm{k}_s}^{p\bt{k}_p}{\hat{a}_{p\bt{k}_p}^{\dagger}\hat{a}_{q\bm{k}_q}\hat{a}_{s\bm{k}_s}}}}
			\label{eq-ip-formula}
		\end{aligned}
	\end{equation}
	\begin{equation}
		\begin{aligned}
			\hat{\mathcal{R}}_{EA}(\bm{k})&=\sum_{p}{r^{p\bm{k}}\hat{a}^\dag_{p\bm{k}}} 
			&+ \sum_{\substack{\bt{k}_p, \bm{k}_q \\ \bm{k}_s}}^{\prime}{\sum_{p, q, s}{r_{s\bm{k}_s}^{p\bt{k}_p q\bm{k}_q}{\hat{a}_{p\bt{k}_p}^{\dagger}\hat{a}_{q\bm{k}_q}^\dag \hat{a}_{s\bm{k}_s}}}}
			\label{eq-ea-formula}
		\end{aligned}
	\end{equation}
	where the $r$-coefficients of the operator basis are parameters to be determined. Again, crystal momentum conservation rule $\bt{k}_{p} \mp \bm{k}_{q} - \bt{k}_{s} \pm \bt{k} =\bm{G}_m$ should be satisfied\cite{McCSunCha17, DittIzs19, WangBer20, GalloHum21, HamesJoh16}. 
	
	In practical EOM calculations, a reference state $| \Psi \rangle$ is chosen. Then, eigenvalues of Eq. (\ref{eq-eom2}) can be obtained from stationary values of the following functional 
	\begin{equation}
		\label{eom-func1}
		\Delta E_{x} = \frac{\langle \Psi | \hat{\mathcal{R}}^{\dagger} [\hat{H}, \hat{\mathcal{R}}] | \Psi \rangle} {\langle \Psi | \hat{\mathcal{R}}^{\dagger} \hat{\mathcal{R}} | \Psi \rangle}
	\end{equation}
	for any reference state  $| \Psi \rangle$ that has a nonzero overlap with the exact ground state $| \Psi_{0} \rangle$. Notice that, if the operator basis to represent $\hat{\mathcal{R}}$ is incomplete, the accuracy of the calculated excitation energy depends on the choice of reference state  $| \Psi \rangle$.
	In the Liouvillian superoperator formalism, Eq. (\ref{eom-func1}) corresponds to a simple metric defined in the operator space $(\hat{\rho_i}|\hat{\rho_j})=\langle \Psi | \hat{\rho_i}^{\dagger} \hat{\rho_j} | \Psi \rangle$ \cite{szekeres2001on}.
	
	Sometimes, it is more convenient to choose the commutator metric $(\hat{\rho_i}|\hat{\rho_j})=\langle \Psi | [\hat{\rho_i}^{\dagger}, \hat{\rho_j}]_+ | \Psi \rangle$. Then, the corresponding EOM functional becomes
	\begin{equation}
		\label{eom-func2}
		\Delta E_{x} = \frac{\langle \Psi | [ \hat{\mathcal{R}}^{\dagger}, [\hat{H}, \hat{\mathcal{R}}] ]_{+}| \Psi \rangle} {\langle \Psi | [ \hat{\mathcal{R}}^{\dagger},  \hat{\mathcal{R}} ]_{+}| \Psi \rangle}.
	\end{equation}
	Such an expression can lower the rank of density matrices involved in classical calculations\cite{EOM-book, szekeres2001on} and reduce the number of Pauli terms in quantum simulations. However, it is not equivalent to the simple metric functional Eq. (\ref{eom-func1}) for a practical ans\"atze of $\hat{\mathcal{R}}$ such as those specified in Eqs. (\ref{eq-ip-formula}) and (\ref{eq-ea-formula}). If the so-called killer condition $\hat{\mathcal{R}}^{\dagger} | \Psi \rangle = 0$ (it means that the ground state cannot be de-excited) is satisfied, the simple and commutator metric expressions becomes equivalent. 
	
	In band structure calculations with complex wave function involved, another important issue is if the excitation energy will be a real number. With such a consideration, a symmetric double commutator form of the EOM functional has been introduced\cite{EOM-book}
	\begin{equation}
		\label{eom-func3}
		\Delta E_{x} = \frac{\langle \Psi | [ \hat{\mathcal{R}}^{\dagger}, \hat{H}, \hat{\mathcal{R}} ]_{+}| \Psi \rangle} {\langle \Psi | [ \hat{\mathcal{R}}^{\dagger}, \hat{\mathcal{R}} ]_{+}| \Psi \rangle}
	\end{equation}
	where $[\hat{\mathcal{R}}^{\dagger}, \hat{H}, \hat{\mathcal{R}} ]_{+} = \frac{1}{2} ([[\hat{\mathcal{R}}^{\dagger}, \hat{H}], \hat{\mathcal{R}}]_{+} + [\hat{\mathcal{R}}^{\dagger}, [\hat{H}, \hat{\mathcal{R}}]]_{+})$. This EOM functional guarantees real-valued excitation energies.  A similar expression has been implemented in the qEOM method to calculate the electron-excitation (EE) energies.\cite{PauAbhChu19} If $\hat{\mathcal{R}} = | \Psi_{x} \rangle \langle  \Psi_0 |$, Eq. (\ref{eom-func3}) can be obtained directly from Eq. (\ref{eq-eom2}). For a practical ans\"atze of $\hat{\mathcal{R}}$, Eq. (\ref{eom-func3}) can be obtained from Eq. (\ref{eom-func2}) if the exact ground state $| \Psi_{0} \rangle$ is used as the reference state.
	
	Since in practical calculations  $\hat{\mathcal{R}}$ is typically truncated and $| \Psi \rangle$ is not the exact ground state, a direct use of Eq. (\ref{eom-func3}) may lead to significant errors as will be demonstrated in Si and diamond band structure calculations. To solve this problem, we adopt the projected excitation operator formalism \cite{szekeres2001on} for both EOM-IP and EOM-EA
	\begin{equation}
		\label{eom-func3-proj}
		\hat{\tilde{\mathcal{R}}}_{IP(EA)} =  \hat{\mathcal{R}}_{IP(EA)} |\Psi \rangle \langle \Psi |
	\end{equation}
	Since $\hat{\tilde{\mathcal{R}}} | \Psi \rangle= \hat{\mathcal{R}} | \Psi \rangle $, we can substituting $\hat{\mathcal{R}}$ in EOM functionals with $\hat{\tilde{\mathcal{R}}}$. Then, the killer condition becomes satisfied for any reference wavefunction. More importantly, the IP or EA energies obtained from Eq. (\ref{eom-func3}) become equivalent to those from Eqs. (\ref{eom-func1}) and (\ref{eom-func2}).
	
	The problem of such a projected operator formalism in quantum simulation is that the projector $|\Psi \rangle \langle \Psi |$ is difficult to be directly implemented in a quantum circuit. \hl{Therefore, a simplified working equation is obtained from either Eqs. (\ref{eom-func1}), (\ref{eom-func2}), or (\ref{eom-func3}), which is given as}
	\begin{equation}
		\label{eom-func3-qse}
		\begin{aligned}
			\Delta E_{x} &= \frac{\langle \Psi | [\hat{\tilde{\mathcal{R}}}^{\dagger}, \hat{H}, \hat{\tilde{\mathcal{R}}} ]_{+}| \Psi \rangle} {\langle \Psi | [ \hat{\tilde{\mathcal{R}}}^{\dagger}, \hat{\tilde{\mathcal{R}}} ]_{+}| \Psi \rangle} \\
			&= \frac{\langle \Psi | \hat{\mathcal{R}}^{\dagger} \hat{H} \hat{\mathcal{R}} | \Psi \rangle} {\langle \Psi | \hat{\mathcal{R}}^{\dagger}  \hat{\mathcal{R}} | \Psi \rangle} - \langle \Psi | \hat{H} | \Psi \rangle 
		\end{aligned}
	\end{equation}
	In contrast to Eqs. (\ref{eom-func1}) and (\ref{eom-func2}), this working equation always leads to real excitation energies. 
	
	The second term of the working equation (\ref{eom-func3-qse}) is a constant and the first term can be converted into a generalized Hermitian eigenvalue problem in a QSE\cite{Mcclean17qse} style
	\begin{equation}
		\bm{H} \bm{C} = \bm{S} \bm{C} \bm{E}
		\label{eq-qse-gegv}
	\end{equation}
	The matrix elements of $\bm{H}$ and $\bm{S}$ can be evaluated as
	\begin{equation}
		\label{eom-func3-proj-mat}
		\begin{aligned}
			&H_{uv} = \langle \Psi_{C} | \hat{\rho}_{u}^{\dagger} \hat{H} \hat{\rho}_{v} | \Psi_{C} \rangle \\
			&S_{uv} = \langle \Psi_{C} | \hat{\rho}_{u}^{\dagger} \hat{\rho}_{v} | \Psi_{C} \rangle \\
		\end{aligned}
	\end{equation}
	where $ \hat{\rho}_{i}$ stands for Fermion excitation terms in Eqs. (\ref{eq-ip-formula}) and (\ref{eq-ea-formula}), $|\Psi_{C} \rangle$ is the ADAPT-C ground-state wave function. Calculation of these matrix elements can take advantage of a quantum computer, while eigenvalues and eigenvectors of Eq.(\ref{eq-qse-gegv}) can be calculated on a classical computer.
	
	Among all the eigenstates, only the quasiparticle states should be considered for band structure. Therefore, we denote $\vec{R}_{1}=\{r_{p\bm{k}}, r^{p\bm{k}}\}$ and define the quasiparticle weight ($QPWT$) as
	\begin{equation}
		QPWT=\norm{\vec{R}_{1}}_2
	\end{equation}
	For example, the $QPWT$ for EOM-IP takes the form
	\begin{equation}
		QPWT_{IP}=\sqrt{\sum_{p}{\abs{r_{p\bm{k}}}^{2}}}.
	\end{equation}
	where $\{r_{p\bm{k}}\}$ are coefficients of single ionization operators $\{\hat{a}_{p\bm{k}}\}$.
	To construct the band structure, we find excitation energies with a large $QPWT$, which indicates the corresponding excitation is dominated by single electron addition (for EOM-EA) or removal (for EOM-IP). The protocol described above for band structure calculation is named EOM-ADAPT-C.
	
	Notice that, in classical EOM-CC calculations, a similarity transformation technique is widely used, where a transformed Hamiltonian $\hat{\tilde{H}} = e^{-\hat{T}} \hat{H} e^{\hat{T}}$ is considered with the Hartree-Fock wave function $|\Psi_{HF}\rangle$ as the reference state\cite{eom-3, Lev2004}. Such a scheme guarantees that the killer condition is satisfied $\hat{\mathcal{R}}^{\dagger} |\Psi_{HF}\rangle \equiv 0$, although it is not due to the orthogonality between excited and ground states as it suppose to be. The similarity-transformed Hamiltonian has the same excitation energy spectrum as the bare Hamiltonian. More importantly, due to the commutation relationship $[\hat{\mathcal{R}}, e^{\pm \hat{T}}]\equiv0$, we also have  $[\hat{\tilde{H}}, \hat{\mathcal{R}}] | \Psi_{HF} \rangle = \Delta E_{x} \hat{\mathcal{R}} | \Psi_{HF} \rangle$, which means that $\hat{\mathcal{R}}$ is the same for the bare and transformed Hamiltonians. \hl{In the UCC or ADAPT case with anti-Hermitian operators $\{\hat{\tau}_{i}\}$ introduced, since $[\hat{\mathcal{R}}, e^{ \hat{\tau}}]\neq0$, $\hat{\mathcal{R}}$ for the bare and transformed Hamiltonians are different. Such a difference is consistent with the intermediate state representation\cite{isr}, where the so-called self-consistent excitation operators $e^{T-T^{\dagger}} \hat{\mathcal{R}} e^{T^{\dagger}-T} $ is applied on the UCC or ADAPT ground state and the similarity-transformed Hamiltonian appears in the working equation. As a result of this difference, the identification of quasiparticle state will become more difficult. At the same time, estimation of the matrix element $\langle \Psi |\hat{\rho}_{u}^{\dagger} e^{T^{\dagger}-T} \hat{H} e^{T-T^{\dagger}} \hat{\rho}_{v}  | \Psi \rangle$ for the similarity-transformed Hamiltonian leads to a more complicated quantum circuit.} Therefore, the similarity transformation technique is not used in this study.
	
	\textbf{Numerical results.}	We first use a one dimensional hydrogen chain system to test the accuracy of ADAPT-C. Then, band structures of diamond and silicon are calculated using the EOM-ADAPT-C algorithm. All calculations are performed using an in-house developed code interfaced with several open-source software packages. One- and two-electron coefficients in the Hamiltonian in Eq. (\ref{eq-ham-pbc}) are calculated with PySCF\cite{pyscf}. The mapping from Fermion operators to Pauli operators are performed using the Jordan-Wigner transformation implemented in the OpenFermion code\cite{openfermion}. In ADAPT-C optimization, the limited-memory Broyden-Fletcher-Goldfarb-Shanno (L-BFGS-B) algorithm in SciPy\cite{scipy} is used. The GTH-SVZ basis set is used together with the GTH pseudopotential. Full configuration interaction (FCI) results are obtained by direct diagonalization of the system Hamiltonian in the qubit space. The convergence threshold of residual gradient $\varepsilon$ in Eq.(\ref{eq-adapt-convg}) is taken to be $1\times10^{-3}$ Hartree in both potential energy surface and band structure calculations.
	
	
	A unit cell with two hydrogen atoms are used to build the one-dimensional hydrogen chain. A $1\times1\times4$ $k$-point grid is used for Brillouin zone sampling. We use the notation ADAPT\{$X$\}, ADAPT-C\{$X$\} and ADAPT-K2G\{$X$\} to represent different types of ADAPT-VQE calculations where $X\in$ \{SD,GSD\}. For example, ADAPT-C\{GSD\} stands for an ADAPT-C calculation with general single and double excitation cluster operators. 
	
	\begin{figure*}[tb]
		\subfigure[]
		{
			\includegraphics[width=0.45\textwidth]{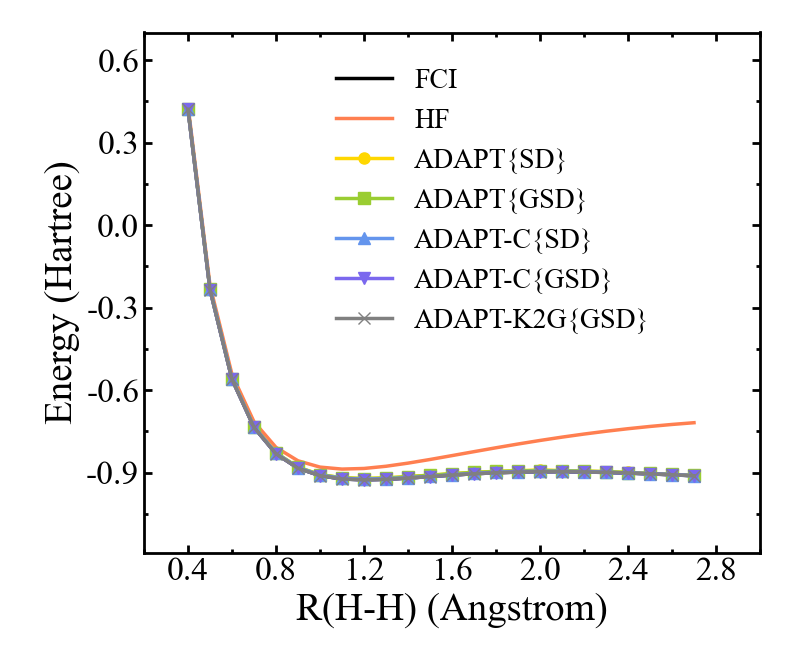}
		}
		\subfigure[]
		{
			\includegraphics[width=0.45\textwidth]{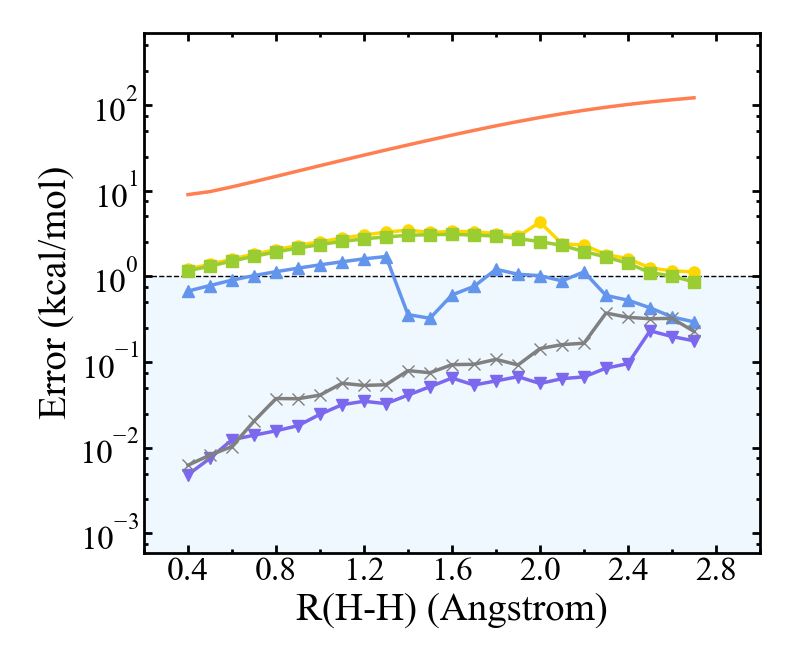}
		}
		\caption{\label{fig1}  (a) Ground-state potential energy surface of one-dimensional hydrogen chain calculated with different methods. (b) Absolute energy error compared with the FCI result. The chemical accuracy (1 kcal/mol) is marked by a dashed line.}
	\end{figure*}
	
	As shown in Figure \ref{fig1}, for both ADAPT and ADAPT-C, the generalized \{GSD\} gives better results than \{SD\}, which is in consistent with previous UCC results\cite{LiuWan20, LeeHugHea19}. More importantly, the ADAPT-C ans\"atze significantly improves the accuracy over ADAPT by 1 to 3 orders of magnitude, since the former eliminating the residual error introduced by complex orbitals.  In the whole range of H-H distance from 0.40 to 2.70 \r{A}ngstrom, the chemical accuracy is reached with ADAPT-C\{GSD\} and the mean absolute error (MAE) throughout the potential energy curve is only 0.06 kcal/mol, which is much smaller than that of ADAPT\{GSD\} (Table \ref{table1}). 
	
	\begin{table*}[bt]
		\centering
		\caption{\label{table1}Mean absolute error (MAE) and maximum absolute error (MaxAE) in ground-state potential energy surface of one-dimensional hydrogen chain in kcal/mol. }
		\begin{tabulary}{1.0\textwidth}{CCCCCCC}
			\hline
			\hline
			&HF&ADAPT \{SD\}&ADAPT \{GSD\}&ADAPT-C\{SD\}&ADAPT-C\{GSD\}&ADAPT-K2G\{GSD\} \\ 
			\hline
			MAE&51.57&2.39&2.13&0.89&0.06&0.12 \\
			MaxAE&121.11&4.28&3.09&1.70&0.23&0.37 \\ 
			\hline
			\hline
		\end{tabulary}
	\end{table*}

	While the accuracy of the K2G transformation technique which we proposed previously\cite{LiuWan20} is at the same order of magnitude as ADAPT-C, the required computational resource is increased roughly by a factor of $N_{k}$, where $N_{k}$ is the total number of $k$-points.  
	The numbers of terms in the Hamiltonian and the size of the operator pool scale as $O(N_{k}^{3} N_{o}^{4} )$ for ADAPT and ADAPT-C, while for the K2G transformation technique this will be increased to $O(N_{k}^{4} N_{o}^{4} )$, where $N_{o}$ is the number of orbitals in a unit cell. The difference becomes even more significant if higher excitation operators such as triples and quadruples are included. Therefore, the ADAPT-C ans\"atze provides an accurate and efficient solution to periodic system ground-state energy calculation.	
	
	
	Band structures of diamond and silicon are calculated at the experimental  lattice constants ($a=5.431$ \r{A} for silicon and $a=3.567$ \r{A} for diamond). At each specific $k$-point $\bm{k}$, a two-step process is executed. (1) \hl{Perform an ADAPT-C\{GSD\} calculation to obtain the ground-state wave function $|\Psi_{C} \rangle$ with a $1\times 1 \times 1$ $k$-point sampling grid centered at $\bm{k}$.} (2) Perform  EOM-ADAPT-C calculations with $\hat{\tilde{\mathcal{R}}}_{IP} | \Psi_{C} \rangle$ and $\hat{\tilde{\mathcal{R}}}_{EA} | \Psi_{C} \rangle$ as the target states. Solve the generalized eigenvalue problem defined in Eq.(\ref{eom-func3-proj-mat}) and compare $QPWT$ to obtain the IP and EA energy spectrum at $\bm{k}$.
	
	\begin{figure}[tb]
		\subfigure[]
		{
			\includegraphics[width=0.45\linewidth]{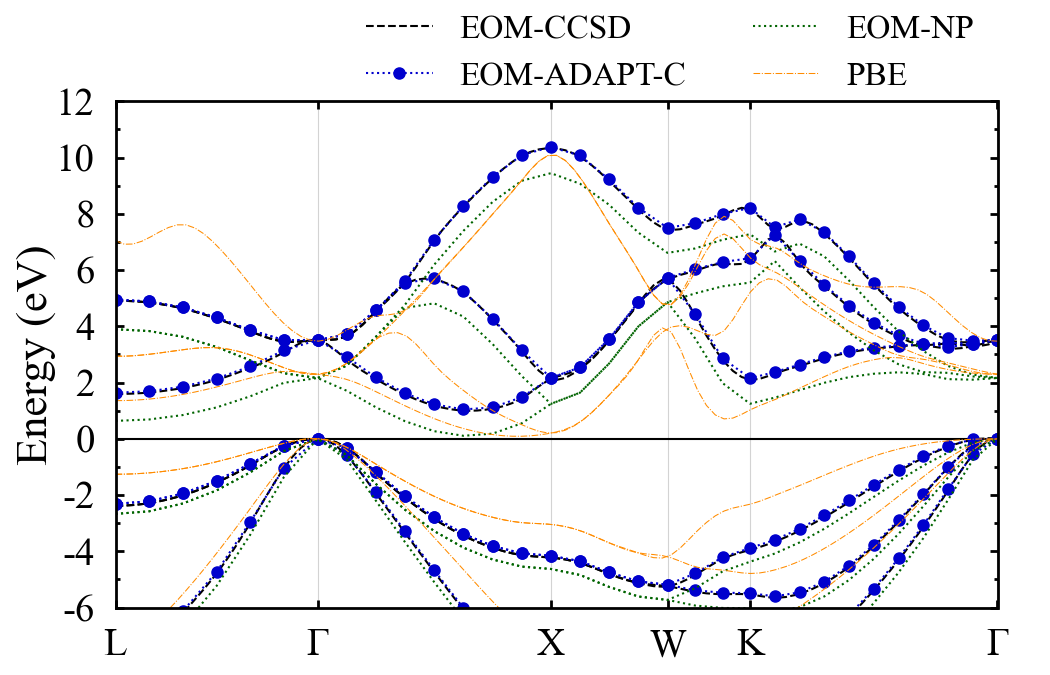}
		}
		\subfigure[]
		{
			\includegraphics[width=0.45\linewidth]{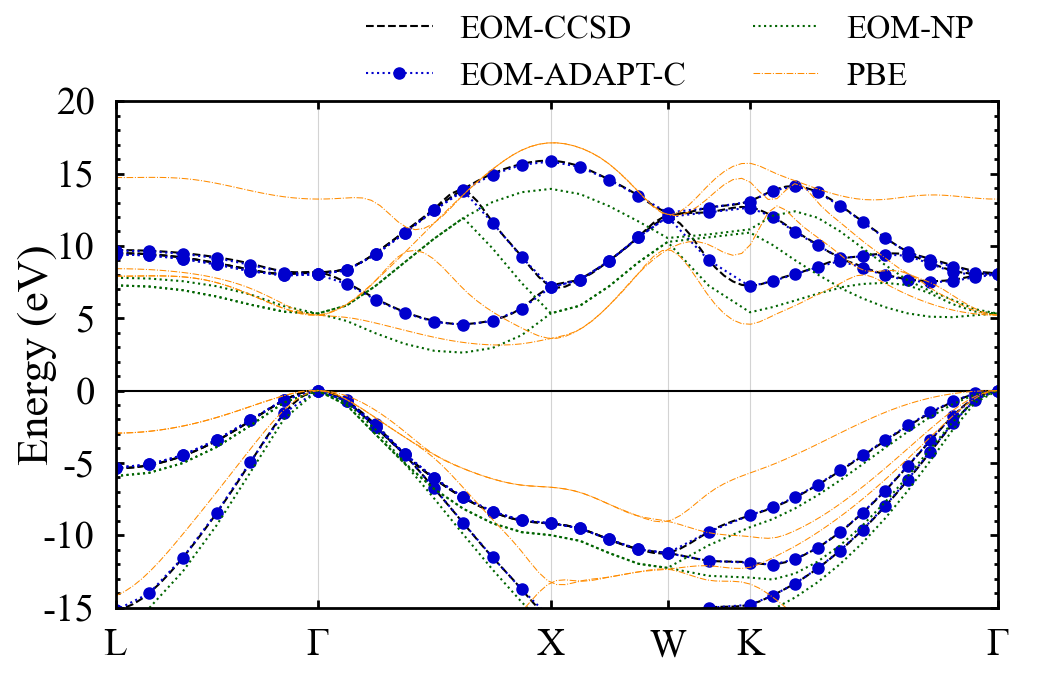}
		}
		\caption{Band structures of (a) silicon and (b) diamond calculated using different methods.}
		\label{fig2s}
	\end{figure}

	As shown in Figure \ref{fig2s}, the band structures calculated with EOM-ADAPT-C agree well with classical EOM-CCSD results. The average deviation is within 0.05 eV. If DFT at the generalized gradient approximation PBE level\cite{pbe} is used to calculate the band structure with the same $k$-point scheme,  a significant deduction of the band gap is observed as expected. On the other hand, if Eq. (\ref{eom-func3}) is used directly without adopting the projected operator Eq. (\ref{eom-func3-proj}), the so-called EOM-NP result is significantly deviated from the EOM-ADAPT-C result. For example, the band gap of Si is changed from 1.04 to 0.11 eV.
	
	\hl{Notice that noise in quantum device should be considered for real applications. Therefore, a noise-included EOM-ADAPT-C simulation to calculate the IP and EA energies of a one-dimensional hydrogen chain is performed. As shown in Figure S1, noise can indeed introduce non-negligible errors in the EOM-ADAPT-C algorithm. These errors can also be significantly reduced by applying error mitigation techniques as for other quantum algorithms. } 
	
	\hl{Relatively small $k$-grids and the minimal basis set are used in this study, which are only suitable for algorithm demonstration and they are not expected to predict accurate electronic structures. Even if clever Brillouin zone sampling technique and accurate Wannier\cite{wannier-2012} basis functions can be used and, at the same time, the requirement to use a high-order truncation in the wave function and excitation operator ans\"atze for strongly correlated systems is not considered, classical simulation of EOM-ADAPT-C algorithm is still strongly restricted by the computational resource cost which grows exponentially. Actually, running an EOM-ADAPT-C simulation on a classical computer is even more expensive than a classical EOM-CC calculation. For example, if the DZVP basis set with a $3\times3\times3$ $k$-point grid is used for C and Si band structure calculations as in the previous classical EOM-CCSD study,\cite{McCSunCha17} a total number of 1404 qubits will be required, leading to a memory consumption of approximately $2^{1404}\times16$ bytes in a brute-force simulation, which is far beyond the capacity of any existing supercomputer.}
	
	However, if the EOM-ADAPT-C algorithm is run on a quantum computer, the expensive construction of Eq.(28) can be realized in parallel via a polynomial number of measurements on a quantum processor. Therefore, the EOM-ADAPT-C algorithm is expected to be a very useful method to obtain highly accurate band structures in the quantum computer era.

	
	In summary, we have proposed the ADAPT-C approach for ground state quantum simulation of solid states and developed a quantum EOM algorithm for quasiparticle band structure calculations. In the ADAPT-C variational ans\"atze, we introduce a set of complementary operators to the original anti-Hermitian operator pool. The complementary operators are targeted at describing the imaginary part of the anti-Hermitian contracted Schr\"odinger equation. Optimizing the energy functional of ADAPT-C wave functions, we reach highly accurate ground-state energy as in the K2G transformation case with a significantly reduced computational cost and improved flexibility. EOM calculations can be performed using the accurate ADAPT-C ground-state wave function as the reference state. A QSE-style EOM equation is derived by adopting projected EOM-IP and EOM-EA operators to make sure that the killer condition is satisfied. As a demonstration, band structures of silicon and diamond are calculated, which is in good agreement with the classical EOM-CCSD result. The EOM-ADAPT-C method developed in this study represents an attractive way for accurate quantum simulation of band structures.

	\begin{acknowledgement}
		This work was partially supported by the NSFC (21825302), by the Fundamental Research Funds for the Central Universities (WK2060000018), and  by the USTC Supercomputing Center. 
	\end{acknowledgement}
	
	\appendix
	\section{Appendices}
	
	\setcounter{figure}{0}
	\renewcommand{\thefigure}{A\arabic{figure}}
	\setcounter{equation}{0}
	\renewcommand{\theequation}{A\arabic{equation}}

	\subsection{A1. Derivation of the EOM-ADAPT-C working equation}
	We starting from Eq. (24) in the main text
	\begin{equation}
		\label{eom-func3-qse}
		\begin{aligned}
			\Delta E_{x} &= \frac{\langle \Psi | [\hat{\tilde{\mathcal{R}}}^{\dagger}, \hat{H}, \hat{\tilde{\mathcal{R}}} ]_{+}| \Psi \rangle} {\langle \Psi | [ \hat{\tilde{\mathcal{R}}}^{\dagger}, \hat{\tilde{\mathcal{R}}} ]_{+}| \Psi \rangle} 
		\end{aligned}
	\end{equation}
	The numerator $N$ and the denominator $D$ of Eq. (\ref{eom-func3-qse}) are
	\begin{equation}
		\begin{aligned}
			N = \langle \Psi | [\hat{\tilde{\mathcal{R}}}^{\dagger}, \hat{H}, \hat{\tilde{\mathcal{R}}}]_{+} | \Psi \rangle \\
			D = \langle \Psi | [\hat{\tilde{\mathcal{R}}}^{\dagger}, \hat{\tilde{\mathcal{R}}}]_{+} | \Psi \rangle
		\end{aligned}
	\end{equation}
	Expand the double commutator for the numerator $N$:
	\begin{equation}
		\begin{aligned}
			2N &= 2 \langle \Psi | [\hat{\tilde{\mathcal{R}}}^{\dagger}, \hat{H}, \hat{\tilde{\mathcal{R}}}]_{+}  |\Psi \rangle \\
			&= \langle \Psi |  [\hat{\tilde{\mathcal{R}}}^{\dagger}, \hat{H}]\hat{\tilde{\mathcal{R}}} + \hat{\tilde{\mathcal{R}}}[\hat{\tilde{\mathcal{R}}}^{\dagger}, \hat{H}] | \Psi \rangle + \langle \Psi | \hat{\tilde{\mathcal{R}}}^{\dagger}[\hat{H}, \hat{\tilde{\mathcal{R}}}] + [\hat{H}, \hat{\tilde{\mathcal{R}}}]\hat{\tilde{\mathcal{R}}}^{\dagger}    | \Psi \rangle \\
			&= \langle \Psi | [\hat{\tilde{\mathcal{R}}}^{\dagger}, \hat{H}]\hat{\tilde{\mathcal{R}}} | \Psi \rangle + \langle \Psi | \hat{\tilde{\mathcal{R}}}^{\dagger}[\hat{H}, \hat{\tilde{\mathcal{R}}}] | \Psi \rangle \\
			&= 2 \langle \Psi | \hat{\tilde{\mathcal{R}}}^{\dagger} \hat{H} \hat{\tilde{\mathcal{R}}} | \Psi \rangle - \langle \Psi | \hat{\tilde{\mathcal{R}}}^{\dagger} \hat{\tilde{\mathcal{R}}} \hat{H} | \Psi \rangle - \langle \Psi | \hat{H} \hat{\tilde{\mathcal{R}}}^{\dagger} \hat{\tilde{\mathcal{R}}} | \Psi \rangle
		\end{aligned}
	\end{equation}
	where,  the killer condition that $\hat{\tilde{\mathcal{R}}}^{\dagger} | \Psi \rangle = \langle \Psi | \hat{\tilde{\mathcal{R}}} = 0$ is applied in the third line. 
	Now substitute $\hat{\tilde{\mathcal{R}}} = \hat{\mathcal{R}} | \Psi \rangle \langle \Psi$ into the equality
	\begin{equation}
		\begin{aligned}
			\langle \Psi | \hat{\tilde{\mathcal{R}}}^{\dagger} \hat{H} \hat{\tilde{\mathcal{R}}} | \Psi \rangle &= 
			\langle \Psi | \Psi \rangle \langle \Psi | \hat{\mathcal{R}}^{\dagger} \hat{H} \hat{\mathcal{R}} | \Psi \rangle \langle \Psi | \Psi \rangle \\
			&= \langle \Psi | \hat{\mathcal{R}}^{\dagger} \hat{H} \hat{\mathcal{R}} | \Psi \rangle
		\end{aligned}
	\end{equation}
	
	\begin{equation}
		\begin{aligned}
			\langle \Psi | \hat{\tilde{\mathcal{R}}}^{\dagger} \hat{\tilde{\mathcal{R}}} \hat{H} | \Psi \rangle &= 
			\langle \Psi | \Psi \rangle \langle \Psi | \hat{\mathcal{R}}^{\dagger} \hat{\mathcal{R}} | \Psi \rangle \langle \Psi | \hat{H} | \Psi \rangle \\ &=
			\langle \Psi | \hat{H} | \Psi \rangle \langle \Psi | \hat{\mathcal{R}}^{\dagger} \hat{\mathcal{R}} | \Psi \rangle
		\end{aligned}
	\end{equation}
	
	\begin{equation}
		\begin{aligned}
			\langle \Psi | \hat{H} \hat{\tilde{\mathcal{R}}}^{\dagger} \hat{\tilde{\mathcal{R}}} | \Psi \rangle &= 
			\langle \Psi | \hat{H} | \Psi \rangle  \langle \Psi | \hat{\mathcal{R}}^{\dagger} \hat{\mathcal{R}} | \Psi \rangle \langle \Psi | \Psi \rangle  \\ &=
			\langle \Psi | \hat{H} | \Psi \rangle \langle \Psi | \hat{\mathcal{R}}^{\dagger} \hat{\mathcal{R}} | \Psi \rangle
		\end{aligned}
	\end{equation}
	The numerator $N$ finally becomes
	\begin{equation}
		\begin{aligned}
			N &= \langle \Psi | [\hat{\tilde{\mathcal{R}}}^{\dagger}, \hat{H}, \hat{\tilde{\mathcal{R}}}]_{+} | \Psi \rangle \\
			&= \langle \Psi | \hat{\mathcal{R}}^{\dagger} \hat{H} \hat{\mathcal{R}}  | \Psi \rangle - \langle \Psi | \hat{H} | \Psi \rangle \langle \Psi | \hat{\mathcal{R}}^{\dagger} \hat{\mathcal{R}} | \Psi \rangle
		\end{aligned}
	\end{equation}
	Similarly, the denominator $D$ can be reduced to
	\begin{equation}
		\begin{aligned}
			D &= \langle \Psi | [\hat{\tilde{\mathcal{R}}}^{\dagger}, \hat{\tilde{\mathcal{R}}}]_{+} | \Psi \rangle \\
			&= \langle \Psi | \hat{\mathcal{R}}^{\dagger} \hat{\mathcal{R}} | \Psi \rangle
		\end{aligned}
	\end{equation}
	Therefore, the final working equation becomes
	\begin{equation}
		\begin{aligned}
			\Delta E_{x} &= \frac{N}{D} = \frac{\langle \Psi | \hat{\mathcal{R}}^{\dagger} \hat{H} \hat{\mathcal{R}}  | \Psi \rangle - \langle \Psi | \hat{H} | \Psi \rangle \langle \Psi | \hat{\mathcal{R}}^{\dagger} \hat{\mathcal{R}} | \Psi \rangle}{\langle \Psi | \hat{\mathcal{R}}^{\dagger} \hat{\mathcal{R}} | \Psi \rangle}\\
			&= \frac{\langle \Psi | \hat{\mathcal{R}}^{\dagger} \hat{H} \hat{\mathcal{R}}  | \Psi \rangle}{\langle \Psi | \hat{\mathcal{R}}^{\dagger} \hat{\mathcal{R}}  | \Psi \rangle} - \langle \Psi | \hat{H} | \Psi \rangle
		\end{aligned}
	\end{equation}
	which is the second or third equality of Eq. (26) in the main text.
	
	\begin{figure*}[p]
		\subfigure[]
		{
			\includegraphics[width=0.45\textwidth]{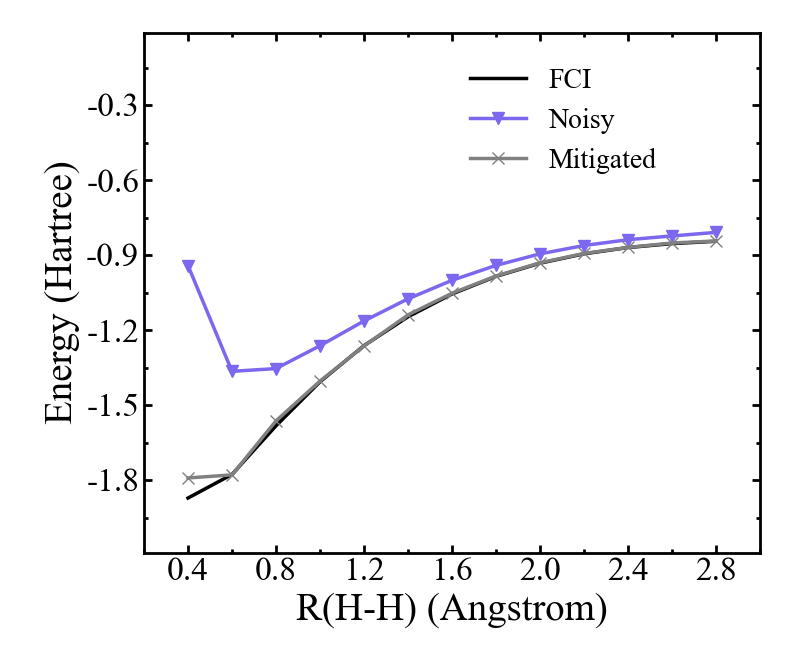}
		}
		\subfigure[]
		{
			\includegraphics[width=0.45\textwidth]{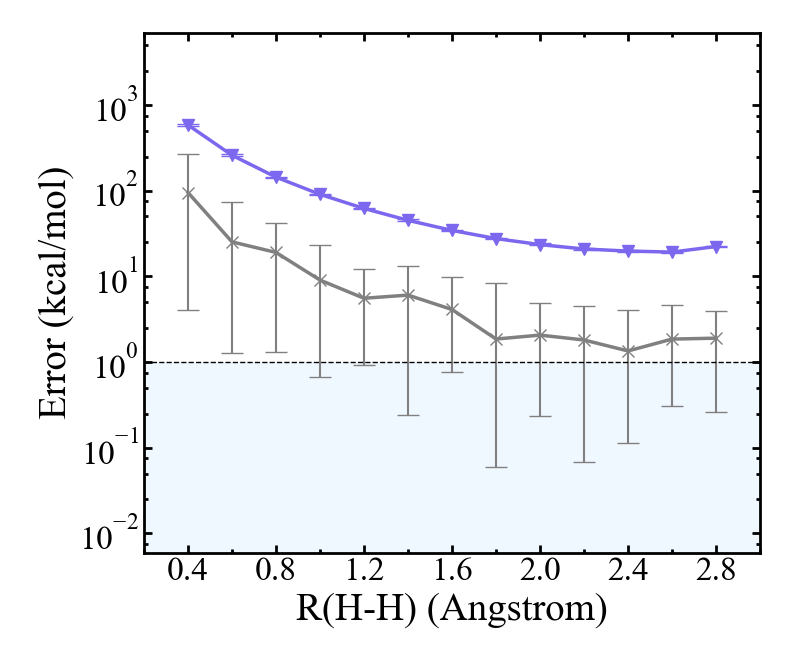}
		}
		\subfigure[]
		{
			\includegraphics[width=0.45\textwidth]{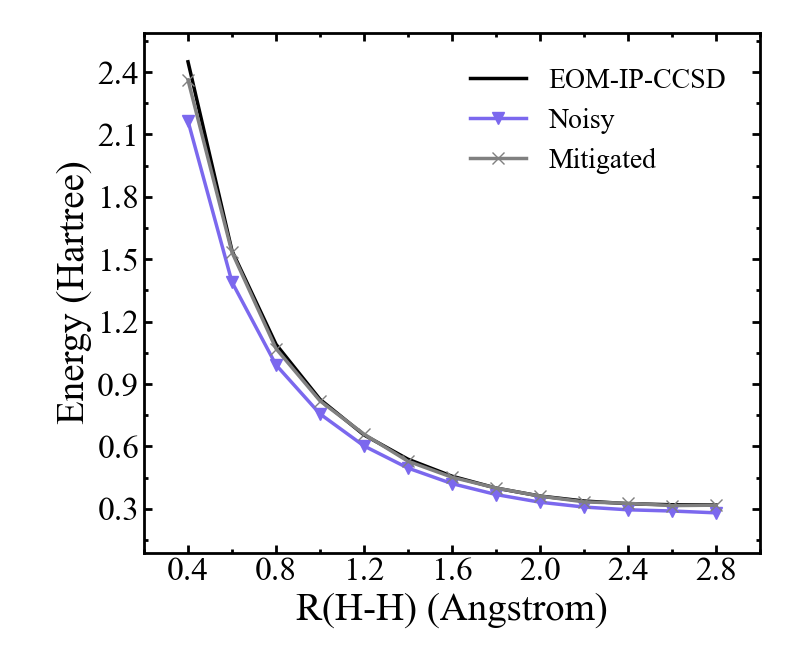}
		}
		\subfigure[]
		{
			\includegraphics[width=0.45\textwidth]{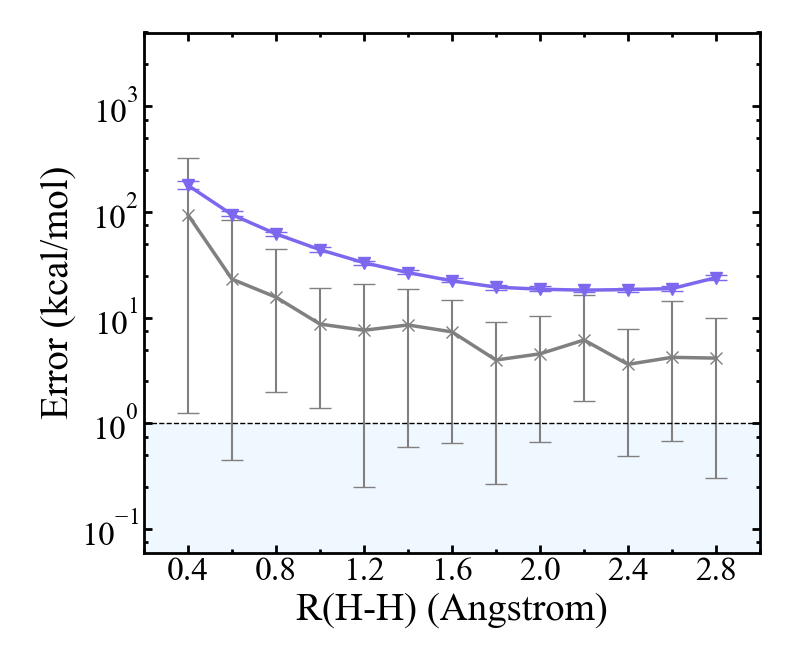}
		}
		\subfigure[]
		{
			\includegraphics[width=0.45\textwidth]{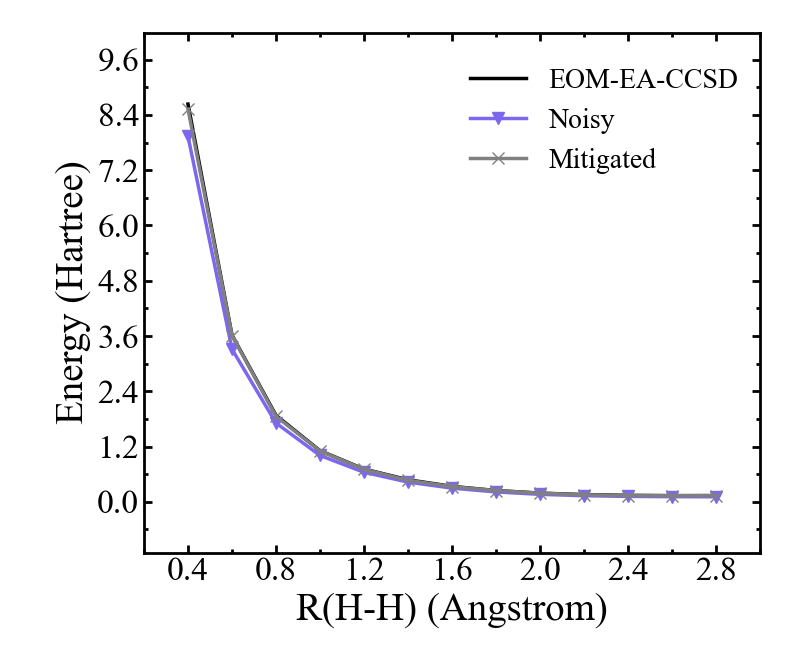}
		}
		\subfigure[]
		{
			\includegraphics[width=0.45\textwidth]{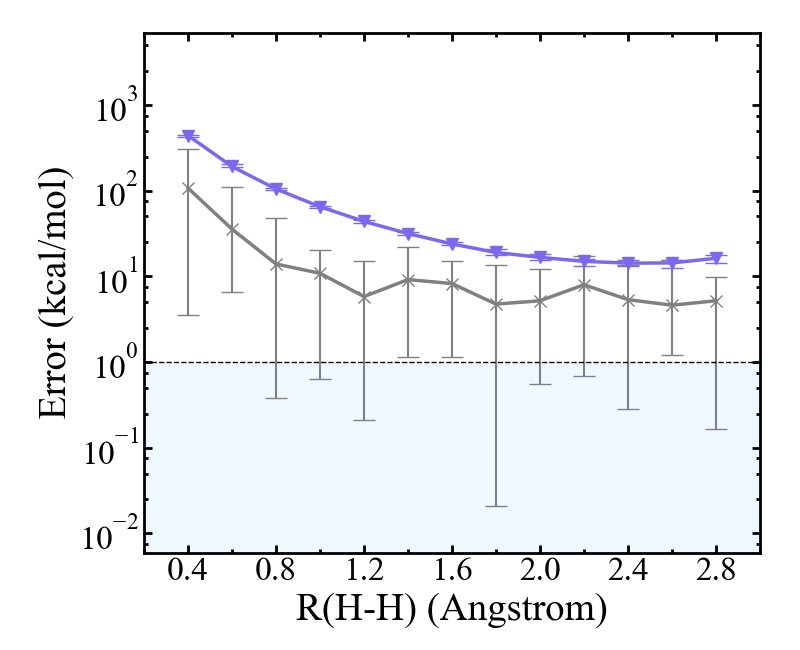}
		}
		\caption{\label{fig-noise}  (a), (b) Ground state potential energy curve and error with respect to FCI results of the one-dimensional hydrogen chain. (c), (d) The lowest IP calculated using EOM-ADAPT-C. The results are compared with classical EOM-IP-CCSD method. (e), (f) Same as (c) or (d), but for the lowest EA. }
	\end{figure*}
	
	\subsection{A2. Noise-included simulation of EOM-ADAPT-C}
	
	Simulation in the presence of noise is performed for the one-dimensional hydrogen chain. The ground-state potential energy curve is calculated using ADAPT-C, and the lowest IP (ionization potential) and EA (electron affinity) states are obtained by subsequent EOM-ADAPT-C calculations. The $k$-point grid used in this test is reduced to $1 \times 1 \times 1$, leading to a 4-qubit quantum circuit. The noise model is implemented by including depolarizing gate errors for all qubits participating in the gates, with the depolarizing parameter $\lambda=0.001$. Error mitigation is performed using the ZNE (zero-noise extrapolation)\cite{zne-2017, zne-2019} technique, with scaling factors $[1.0, 1.25, 1.50]$ and linear extrapolation method. Expectation values are averaged over $2^{17}$ trials, and the experiment is repeatedly carried out for 16 times. All simulations are performed using the Qiskit toolkit\cite{qiskit-package} with the built-in QASM simulator and the Mitiq package\cite{mitiq-package}.

	The results of the calculated ground-state and excitation energies are summarized in Figure.(S\ref{fig-noise}). For the ground state, one double excitation cluster operator is selected by the ADAPT algorithm. The noise introduced by depolarizing gate errors leads to remarkable errors in both the ground-state and the IP or EA energies. A linear extrapolation model is able to reduce the errors significantly.
	
	\bibliographystyle{achemso}
	\bibliography{citations-full.bib}
	
\end{document}